\documentclass{PoS}
\usepackage{epsfig}
\usepackage{bm}
\usepackage{amssymb}
\usepackage{fontenc}
\usepackage{times}
\usepackage{mathptmx}
\usepackage{graphicx}

\title{Complex Langevin for Lattice QCD}

\ShortTitle{Complex Langevin for Lattice QCD}

\author{\speaker{D.~K.~Sinclair}%
         \thanks{This research was supported in part by US Department of Energy
         contract DE-AC02-06CH11357}
         \\
HEP Division, Argonne National Laboratory, 9700 South Cass Avenue, Argonne, 
Illinois 60439, USA\\
E-mail: \email{dks@hep.anl.gov}}

\author{J.~B.~Kogut\\
Department of Energy, Division of High Energy Physics, Washington, DC 20585,
USA\\
and\\
Department of Physics -- TQHN, University of Maryland, 82 Regents Drive, 
College Park, MD 20742, USA\\
        E-mail: \email{jbkogut@umd.edu}}

\abstract{We simulate lattice QCD at finite quark-number chemical potential,
$\mu$, using the complex-Langevin equation (CLE) with gauge-cooling and
adaptive updating to prevent instabilities. The CLE is used because QCD at
finite $\mu$ has a complex fermion determinant which precludes the use of
standard simulation methods based on importance sampling. Since, even when CLE
simulations converge, they are not guaranteed to produce correct results
except under very stringent conditions, which lattice QCD at finite $\mu$
does not obey, we need extensive testing to determine under what conditions it
produces reliable results. We performed simulations at $\beta=6/g^2=5.6$ and
$\beta=5.7$, both at $m=0.025$. For small $\mu$ and $\mu$ large enough to 
produce saturation, measured observables appear to be approaching their correct
values as the coupling is decreased. However, for intermediate $\mu$ values,
these simulations predict a transition from hadronic to nuclear matter at a
$\mu$ which is far too small. Since there is evidence that for CLE simulations
to produce correct results the trajectories should remain close to the $SU(3)$
manifold (at least for small $\mu$), we explore the parameter space to see where
this is true. We find that the distance from this manifold decreases as the
coupling decreases and as the quark mass (in lattice units) decreases, i.e. as
we approach the continuum limit. This indicates that we need to simulate at
smaller couplings and quark masses (requiring larger lattices) to see if these
can produce the correct physics.}

\FullConference{The 36th Annual International Symposium on Lattice Field Theory - LATTICE2018\\
		22-28 July, 2018\\
		Michigan State University, East Lansing, Michigan, USA.}

\begin{document}

\section{Introduction}

QCD at finite quark/baryon-number density describes nuclear matter, the
constituent of the interiors of neutron stars and heavy nuclei. Hot nuclear
matter is produced in relativistic heavy-ion colliders, and was present in the
early universe.

QCD at finite baryon/quark-number density has a sign problem which prevents
direct application of standard lattice simulations that are based on importance
sampling. When finite density is implemented by introducing a quark-number
chemical potential $\mu$, the sign problem manifests itself by making the
fermion determinant complex. This precludes the direct application of standard
lattice QCD simulations based on importance sampling. Since Langevin
simulations are not based on importance sampling, they can be extended to the
case of complex actions
\cite{Parisi:1984cs,Klauder:1983nn,Klauder:1983zm,Klauder:1983sp}. 
For lattice QCD this requires analytically continuing
the gauge fields from $SU(3)$ to $SL(3,C)$. Complex Langevin (CLE) simulations
cannot be guaranteed to produce correct results unless the trajectories are
restricted to a compact domain, the drift term is holomorphic in the fields
and the solutions are ergodic
\cite{Aarts:2009uq,Aarts:2011ax,Nagata:2015uga,Nishimura:2015pba,Nagata:2016vkn,
Aarts:2017vrv,Seiler:2017wvd,Aarts:2017hqp}. 
However, zeroes of the fermion determinant produce poles in the drift term
making it meromorphic not holomorphic in the fields. Thus convergence to the
correct limits cannot be guaranteed. CLE simulations of lattice QCD at finite 
$\mu$ with heavy quarks have been performed by 
\cite{Aarts:2008rr,Aarts:2013uxa,Aarts:2014bwa,Aarts:2016qrv,Langelage:2014vpa}.
CLE simulations with lighter quarks have been performed by
\cite{Sexty:2013ica,Aarts:2014bwa,Fodor:2015doa,Nagata:2016mmh,Scherzer}. For a
good summary of recent work on applying the CLE to lattice QCD at finite $\mu$
with a guide to the literature see \cite{Aarts:2017vrv}.

Our investigations \cite{Sinclair:2017zhn} are aimed at determining whether
the CLE is a viable way of simulating QCD at finite $\mu$, and if so, under
what conditions. We have performed CLE simulations of lattice QCD at zero
temperature and $\mu$s ranging from zero to saturation, at $\beta=6/g^2=5.6$
and $\beta=5.7$, both at $m=0.025$ The weaker coupling shows good agreement
with expectations at small and large $\mu$s, but fails for couplings in the
transition region. The results are compared with those of the phase-quenched
approximation, since random matrix theory suggests that when the CLE fails, it
produces phase-quenched results \cite{Bloch:2017sex}. Other random matrix CLE
simulations seem more optimistic \cite{Nagata:2016alq} In addition full and
phase-quenched simulations are expected to agree at small and large $\mu$.

Since it appears that good results might be obtained with the CLE if the
trajectories remain close to the $SU(3)$ manifold, we are studying how the
unitarity norm, which measures this closeness, depends on quark mass ($m$) and
$\beta$. We find that the average distance from this manifold decreases as $m$
decreases and as the coupling $g$ decreases ($\beta$ increases), i.e. as we
approach the continuum limit.

\section{Complex Langevin Equation for Lattice QCD at finite $\mu$}

If $S(U)$ is the gauge action after integrating out the quark fields, 
The Langevin equation for the evolution of the gauge fields $U$ in Langevin
time $t$ is:
\begin{equation}
-i \left(\frac{d}{dt}U_l\right)U_l^{-1} = -i \frac{\delta}{\delta U_l}S(U)
+\eta_l
\end{equation}
where  $S(U)$ is the gauge action after integrating out the quark fields. $l$
labels the links of the lattice, and $\eta_l=\eta^a_l\lambda^a$. Here
$\lambda_a$ are the Gell-Mann matrices for $SU(3)$ and $\eta^a_l(t)$ are
Gaussian-distributed random numbers normalized so that: \begin{equation}
\langle\eta^a_l(t)\eta^b_{l'}(t')\rangle=\delta^{ab}\delta_{ll'}\delta(t-t')
\end{equation}

The complex-Langevin equation has the same form except that the $U$s are now
in $SL(3,C)$. $S$, now $S(U,\mu)$ is 
\begin{equation}
S(U,\mu) =  \beta\sum_\Box \left\{1-\frac{1}{6}Tr[UUUU+(UUUU)^{-1}]\right\} 
-\frac{N_f}{4}{\rm Tr}\{ln[M(U,\mu)]\}
\end{equation}
where $M(U,\mu)$ is the staggered Dirac operator, backward links are
represented by $U^{-1}$ not $U^\dag$ and we choose to keep the noise term
$\eta$ real. We simulate the time evolution of the gauge fields using a
partial second-order formalism, and stochastic estimators for 
${\rm Tr}\{ln[M]\}$ \cite{Ukawa:1985hr,Fukugita:1986tg,Fukugita:1988qs}

We apply adaptive updating: if the force term becomes too large, $dt$ is
decreased to keep it under control. After each update, we gauge cool
\cite{Seiler:2012wz}, gauge fixing to the gauge which minimizes the unitarity 
norm:
\begin{equation}
F(U) = \frac{1}{4V}\sum_{x,\mu}Tr\left[U^\dag U + (U^\dag U)^{-1} - 2\right]
                                            \ge 0\;.
\end{equation}

We use unimproved staggered quarks.

\section{Zero Temperature Simulations at $\beta=5.6$ and $\beta=5.7$, $m=0.025$}

We perform CLE simulations of 2-flavour lattice QCD at zero temperature at
$\beta=5.6$, $m=0.025$ on a $12^4$ lattice and at $\beta=5.7$, $m=0.025$ on a
$16^4$ lattice from $\mu=0$ up to saturation. For comparison, we perform RHMC
simulations of the phase-quenched approximation over the same parameter range,
since random matrix theory suggests that when the CLE simulations fail they
produce the phase-quenched results.

The phase-quenched approximation is known to undergo a phase transition to a
superfluid phase at $\mu\approx m_\pi/2$. The chiral condensate is constant up
to this transition and decreases beyond it, vanishing at saturation. The
quark-number density is zero up to the transition beyond which it rises up to
saturation, where all states are filled (density=3 in our normalization). At
saturation the quarks decouple and we have a pure gauge theory. For the full
theory one expects the observables to remain at their $\mu=0$ values up to
$\mu\approx m_N/3$ above which they evolve towards saturation. For
$\beta=5.6$, $m=0.025$, $m_\pi/2 \approx 0.21$, $m_N/3 \approx 0.33$
\cite{Bitar:1990cb}, while for $\beta=5.7$, $m=0.025$, $m_\pi/2 \approx
0.194$, $m_N/3 \approx 0.28$ \cite{Brown:1991qw,Schaffer:1992rq}. We typically
run for 2-3 million updates at each $\beta$ and $\mu$.

\begin{figure}[htb]
\vspace*{-0.2in}
\parbox{2.9in}{
\epsfxsize=2.9in                                                       
\epsffile{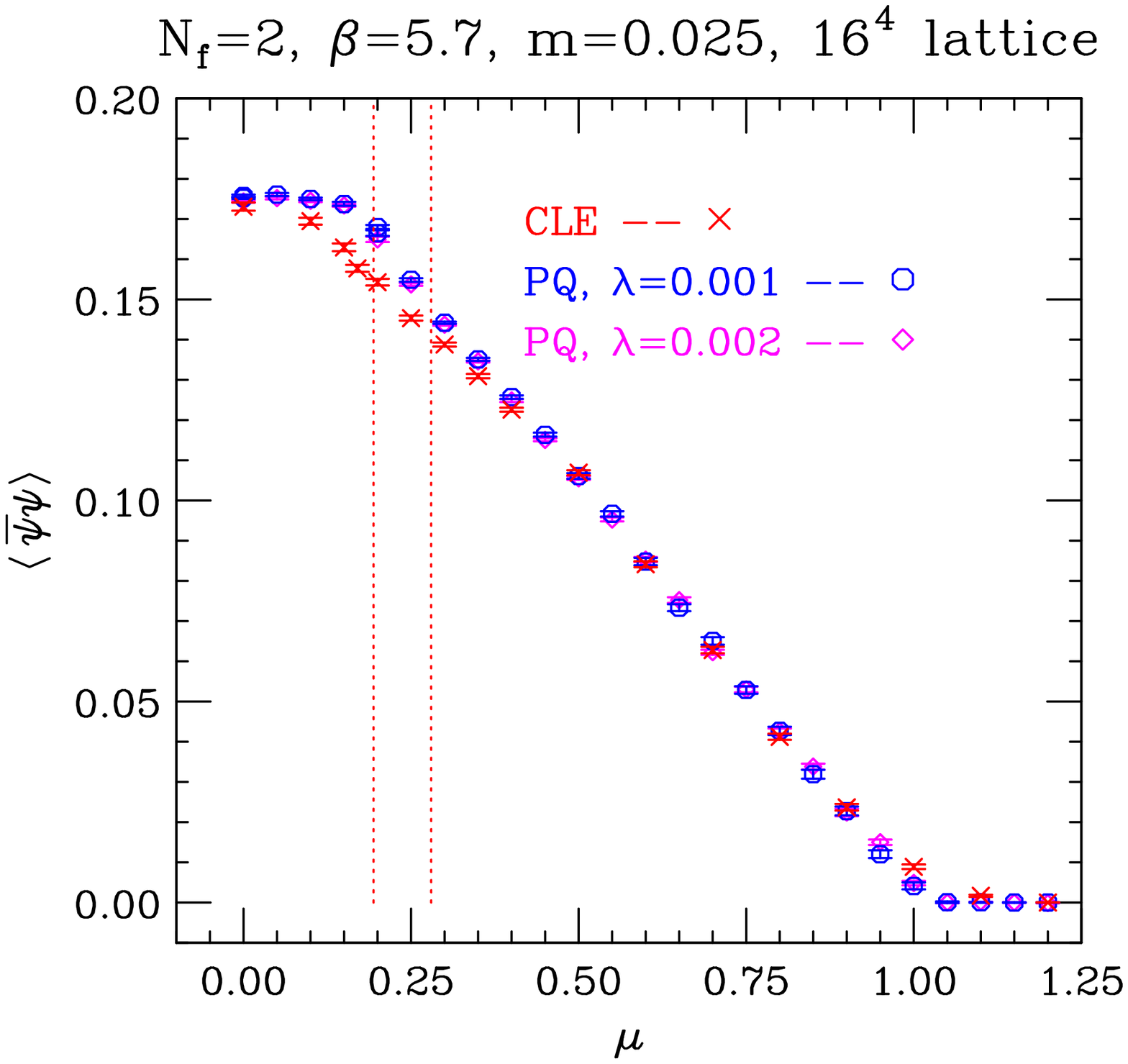}
\caption{Chiral condensate as a function of $\mu$ on a $16^4$ lattice
at $\beta=5.7$ and $m=0.025$.}
\label{fig:pbp}  
}
\parbox{0.2in}{}
\parbox{2.9in}{
\epsfxsize=2.9in
\centerline{\epsffile{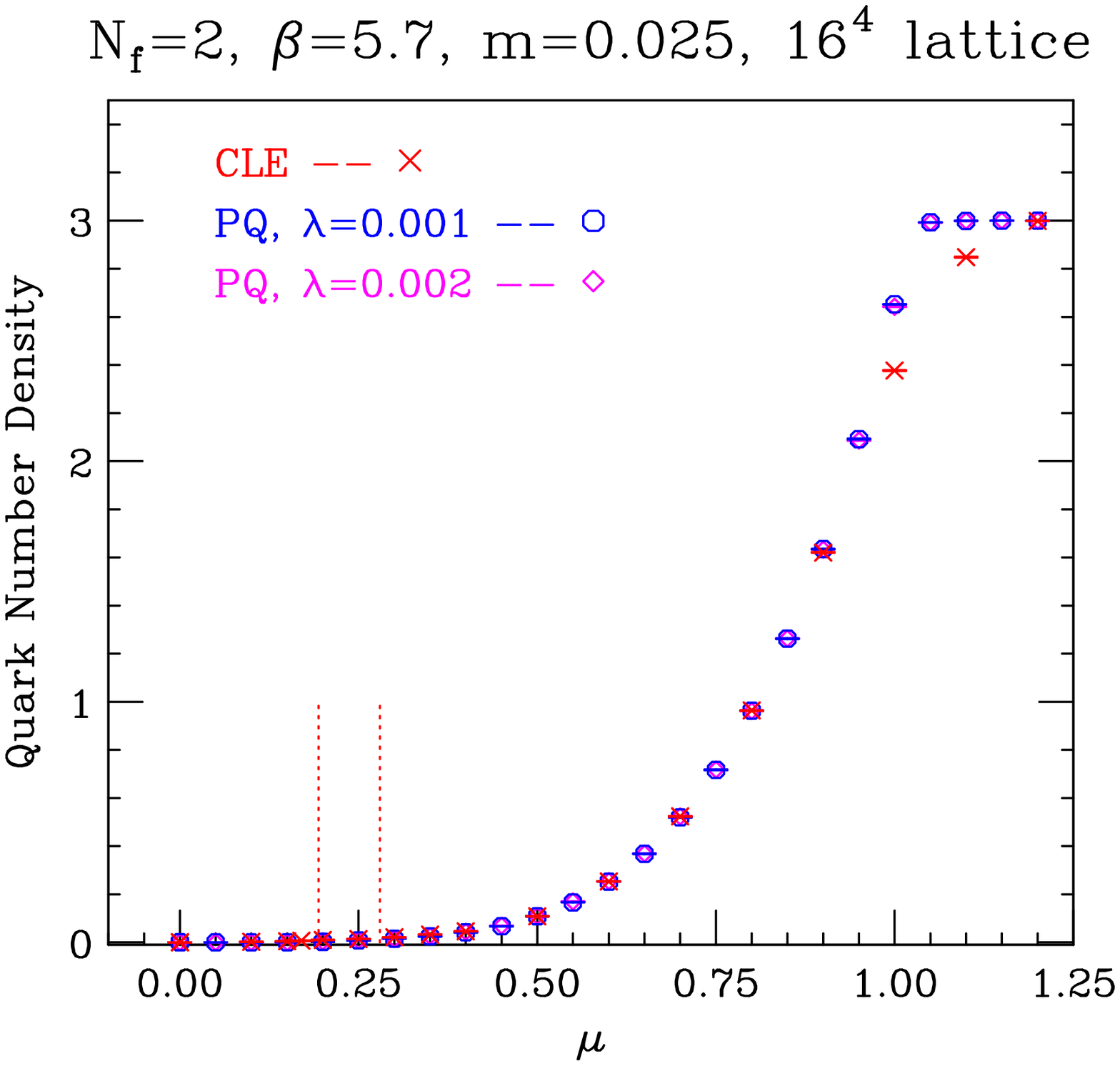}}
\caption{Quark-number density as a function of $\mu$ on a $16^4$ lattice
at $\beta=5.7$ and $m=0.025$.}
\label{fig:qnd}
}
\parbox{2.9in}{ 
\epsfxsize=2.9in
\vspace{0.1in}
\epsffile{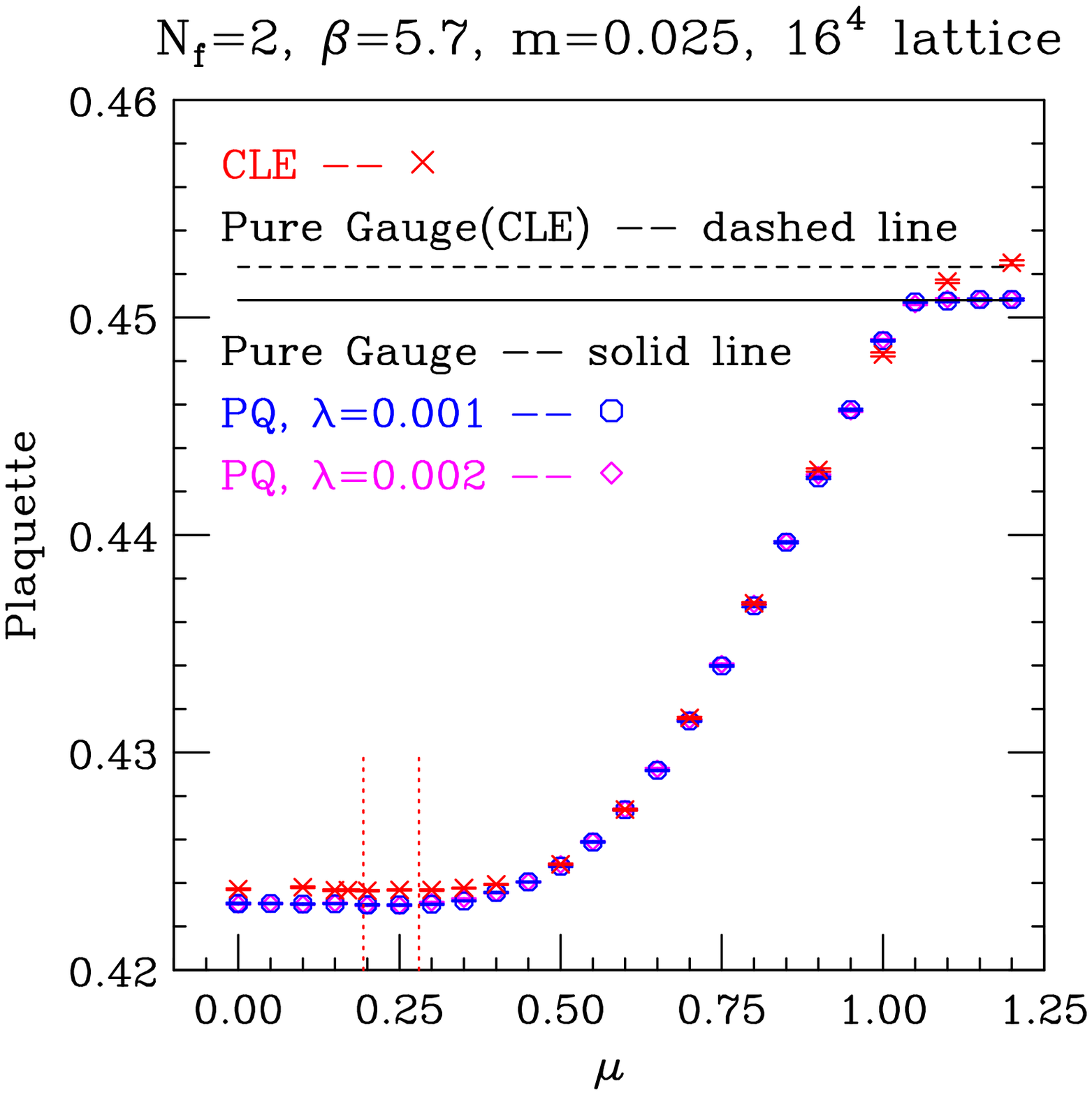}
\caption{Plaquette as a function of $\mu$ on a $16^4$ lattice    
at $\beta=5.7$ and $m=0.025$.} 
\label{fig:plaquette}
}
\parbox{0.2in}{}                                                        
\parbox{2.9in}{                                                              
\epsfxsize=2.9in
\vspace{0.1in}   
\epsffile{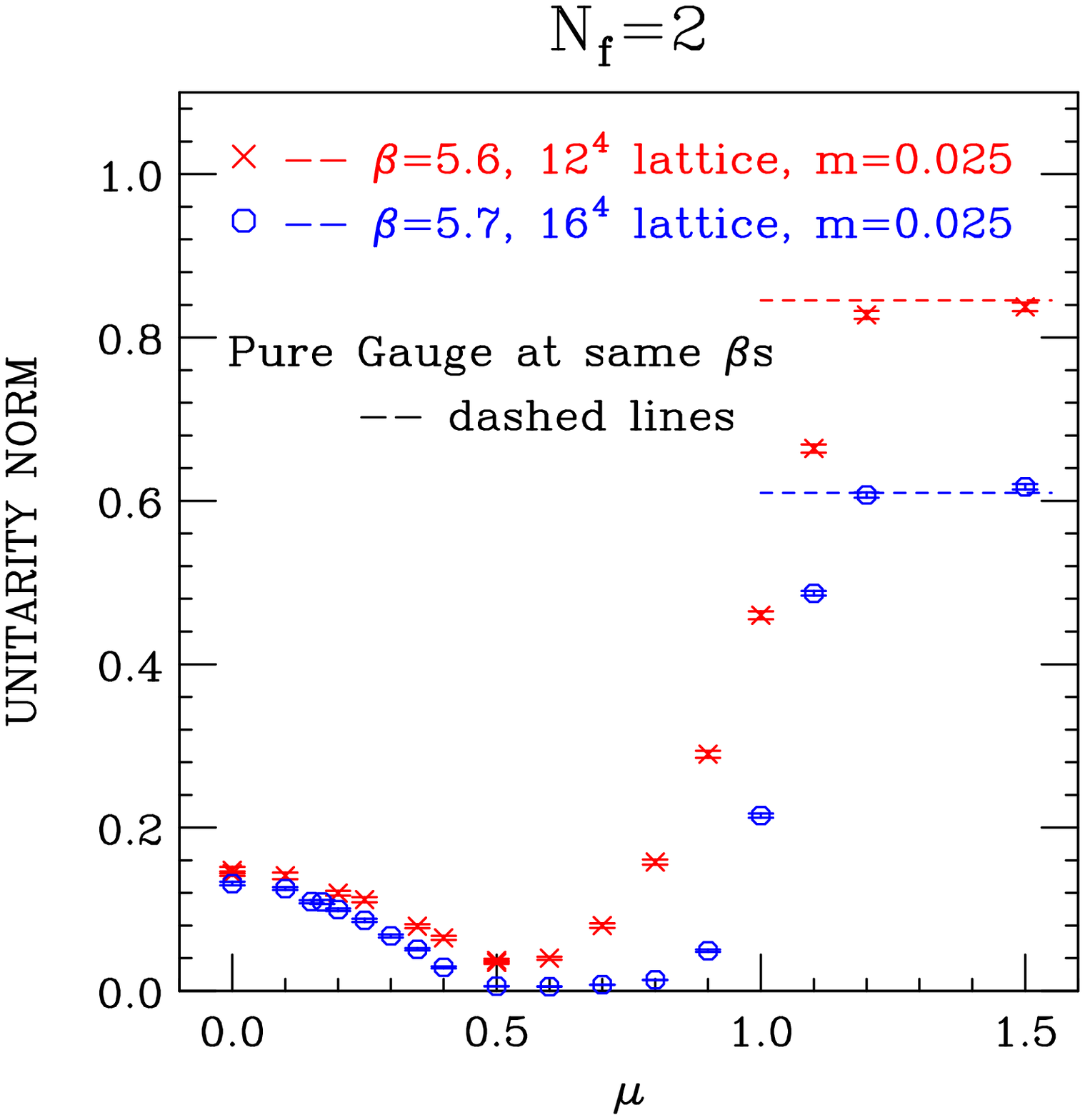}
\caption{Average unitarity norm  as a function of $\mu$ for $\beta=5.6$,
on a $12^4$ lattice -- red, and for  $\beta=5.7$ on a $16^4$ lattice -- blue.} 
\label{fig:unorm}
}
\end{figure}

Figure~\ref{fig:pbp} shows the chiral condensate 
($\langle\bar{\psi}\psi\rangle$), figure~\ref{fig:qnd} the quark-number density
and figure~\ref{fig:plaquette} the plaquettes, as functions of $\mu$ at 
$\beta=5.7$, $m=0.025$ on a $16^4$ lattice. The vertical red dotted lines are
at $\mu=m_\pi/2$ and $\mu=m_N/3$ respectively. For $\mu$ at or near zero and
for $\mu$ at saturation, where the quarks decouple and the gauge fields
exhibit pure gauge dynamics, these observables are in good agreement with
known values, a considerable improvement from $\beta=5.6$. However, for both
$\beta$s, these graphs show a transition from hadronic to nuclear matter at a
$\mu < m_\pi/2$, rather than at $\mu \approx m_N/3$, a result even worse than
the phase-quenched approximation.

The unitarity norm is significantly greater than zero throughout the transition
region for both $\beta$ values, only becoming small around $\mu=0.5$. For
$\beta=5.7$, it remains small through $\mu=0.8$ before increasing towards its
pure gauge value at saturation. We conjecture that this means we can trust
the CLE for $\mu \ge 0.5$.

\section{Dependence of the Unitarity Norm on $m$ and on $\beta=6/g^2$}

The behaviour of the CLE seems to improve if the gauge fields remain close to
the $SU(3)$ manifold, i.e. if the unitarity norm remains small. It is thus
useful to study how the average unitarity norm depends on the simulation
parameters. Therefore we study how this norm depends on $m$ and $\beta$. Since
we have seen that in the small $\mu$ regime ($\mu < 0.5$), where failures of
the CLE first manifest themselves, the unitarity norm decreases as $\mu$
increases from zero, it will suffice to restrict ourselves to $\mu=0$.
\begin{figure}[htb]
\vspace*{-0.2in}
\parbox{2.9in}{
\epsfxsize=2.9in
\epsffile{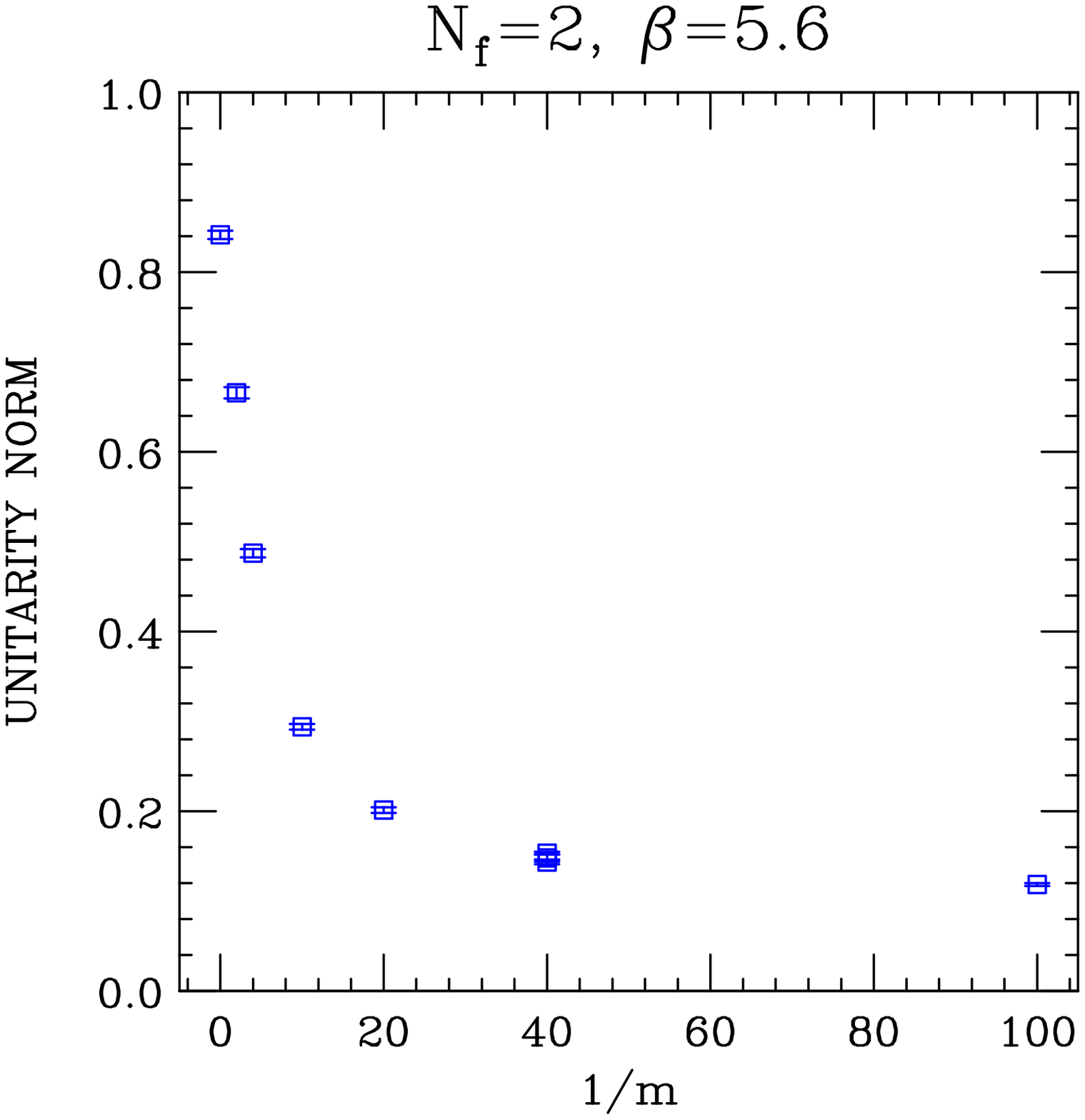}
\caption{Unitarity norm as a function of inverse quark mass at $\beta=5.6$}
\label{fig:unorm(m)}
}
\parbox{0.2in}{}
\parbox{2.9in}{
\epsfxsize=2.9in
\epsffile{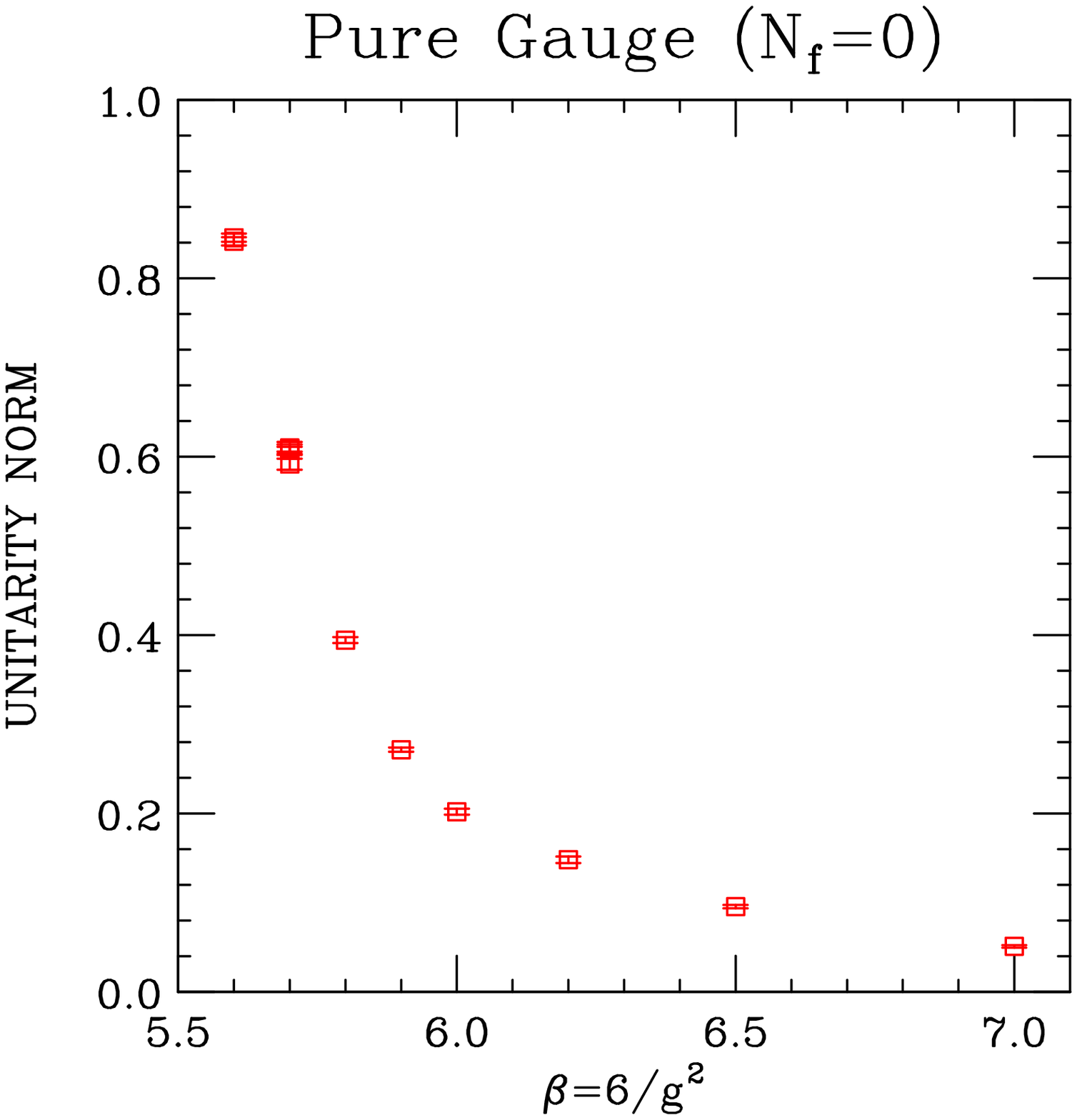}
\caption{Unitarity norm as a function of $\beta=6/g^2$ for pure 
$SU(3)$ gauge theory.}
\label{fig:unorm(beta)}
}
\end{figure}
With $\beta$ fixed at $5.6$ and $\mu=0$, we determine the dependence of the
unitarity norm on the quark mass $m$ for $0.1 \le m \le \infty$. This
dependence is shown in figure~\ref{fig:unorm(m)}. The unitarity norm decreases
as $m$ decreases falling by a factor of about 7 over this range. 

We have already seen that, with $m$ fixed at $0.025$ the unitarity norm
decreases when $\beta$ is increased from $5.6$ to $5.7$. Since the unitarity
norm has its maximum for $m=\infty$, i.e. for pure $SU(3)$ Yang-Mills theory,
we choose to apply the CLE for this mass. Because there are no quarks such
simulations are cheap and can be performed on smaller lattices than for lighter
quarks. In addition, without quarks, the action {\it is} holomorphic in the
fields, so that provided that the regions over which they evolve are strongly
bounded (and simply connected), the CLE should be valid. We have run CLE
simulations for $\beta$s in the range $5.6 \le \beta \le 7.0$. The unitarity
norms decrease as $\beta$ is increased ($g$ is decreased), falling by more
than an order of magnitude over this range. This is shown in
figure~\ref{fig:unorm(beta)}. Except at $\beta=5.6$ the plaquette values are in
excellent agreement with the exact (monte-carlo) results indicating that the
CLE produces correct results. We note that our $\mu=0$ simulations at 
$\beta=5.8$, $m=0.02$ and $\beta=5.9$, $m=0.015$ on a $32^4$ lattice give
further indications that the unitarity norm decreases as $\beta$ is increased 
and $m$ is decreased.

The indications are that the unitarity norm will approach zero as $m$ and $g$
approach zero, i.e. in the continuum limit. This gives us hope that the CLE
might deliver correct results in this limit.

\section{Discussion and Conclusions}

We have simulated 2-flavour lattice QCD at $\beta=5.6$, $m=0.025$ and
$\beta=5.7$, $m=0.025$ from $\mu=0$ to $\mu=1.5$ (saturation) using the CLE
with gauge-cooling and adaptive updating. These runs were performed on lattices
which are (at least for small $\mu$s) at zero temperature. For $\mu$ small, and
for $\mu$ at saturation, the observables appear to approach their physical
values as $\beta$ is increased. At intermediate $\mu$ values the transition
from hadronic to nuclear matter occurs at an unphysically small $\mu$.

It has been observed that CLE simulations are more likely to give the
correct results if the trajectories remain close to the $SU(3)$ manifold. We
study the dependence of the average distance from this manifold on (lattice)
quark mass $m$ and on $\beta=6/g^2$, and find that this decreases as $m$ and
$g$ decrease, i.e. as the continuum limit is approached. This suggests that
we might find correct physics, even in the transition region, for large enough
$\beta$s and small enough $m$s. It remains an open question whether the
continuum limit of CLE simulations will produce the correct physics or, as
suggested by random matrix theories, phase-quenched results.

We have performed some exploratory CLE runs at finite temperatures on $12^3
\times 6$ lattices, with $m=0.025$ However, for 2-flavours and $m=0.025$, we
really need $N_t$ large enough that $\beta=5.6$ is on the low temperature side
of the transition from hadronic/nuclear matter to a quark-gluon plasma, to
have any chance of getting the correct physics. This means we would need $N_t
\ge 12$. This situation is worse than that with 4-flavour QCD 
\cite{Fodor:2015doa}.

Because of the flavour (`taste') breaking for staggered quarks, it is possible
that Wilson quarks would be better suited for CLE simulations of lattice QCD
at finite $\mu$. Other action modifications such as the inclusion of chiral
4-fermion interactions might produce better results.

We should mention methods which are being tried by other researchers to
improve the performance of CLE simulations of QCD at finite $\mu$. These
include modifying the dynamics to include irrelevant terms which keep the
unitarity norms closer to the $SU(3)$ manifold \cite{Attanasio:2018rtq}, and
including additional terms in the action, which improve the behaviour of the
CLE, with coefficients that can be continued to zero afterwards
\cite{Nagata:2018mkb}.

\section*{Acknowledgments}

Our simulations are performed on the Bebop cluster at Argonne's LCRC, Crays
Cori and Edison at NERSC, the Stampede 2 cluster at TACC, the Bridges cluster
at PSC, the Comet cluster at SDSC and Linux PCs belonging to the HEP division
at Argonne.

\end{document}